\documentstyle [12pt] {article}

\newcommand {\be} {\begin{equation}}
\newcommand {\bea} {\begin{eqnarray}}
\newcommand {\ee} {\end{equation}}
\newcommand {\eea} {\end{eqnarray}}
\newcommand {\bi} {\bibitem}
\newcommand {\si} {\sigma}
\newcommand {\al} {\alpha}
\newcommand {\g} {\gamma}
\newcommand {\ep} {\epsilon}
\newcommand {\om} {\omega}
\newcommand {\p} {\psi}
\newcommand {\la} {\lambda}

\newcommand {\ie} {{\it i.~e.}}

\newcommand {\x} {\vec{x}}
\newcommand {\y} {\vec{y}}
\newcommand {\gd} {g^{\dagger}}
\renewcommand{\theequation}{\thesection.\arabic{equation}}

\begin{document}
\title {Exact solution of a one-dimensional continuum percolation model}
\author{ Alon Drory}
\date{ \it Dipartimento di Fisica, Universit\`a {\sl La Sapienza}\\
        Piazzale Aldo Moro 2, Roma 00185, Italy}
\maketitle

\begin{abstract}
I consider a one dimensional system of particles which interact through a
hard core of diameter $\si$ and can connect to each other if they are closer
than a distance $d$. The mean cluster size increases as a function of the
density $\rho$ until it diverges at some critical density, the percolation 
threshold. This system can be mapped onto an off-lattice generalization of the
Potts model which I have called the Potts fluid, and in this way, the mean
cluster size, pair connectedness and percolation probability can be calculated
exactly. The mean cluster size is $S = 2 \exp[ \rho (d -\si)/(1 - \rho \si)] - 
1$ and diverges only at the close packing density $\rho_{cp} = 1 / \si $. This
is confirmed by the behavior of the percolation probability. These results
should help in judging the effectiveness of approximations or simulation
methods before they are applied to higher dimensions.
\end{abstract}


\section{Introduction}
\setcounter{equation}{0} 

One dimensional models have a long history in phase transition studies, going
back to Ising's solution of the model which bears his name. Such models have 
been found to be free of phase transitions except in some singular 
circumstances (such as zero temperature), but nevertheless keep being studied
because they are sometimes exactly solvable. Because of this, these models
can serve as testing grounds, for example, for approximations which are 
then used in other dimensions. In this spirit, I present here the exact 
solution of a continuum percolation model in one dimension. Like its brethren,
this model exhibits a phase transition only in some singular circumstances. 
However, it being exactly solvable may yet make it interesting.

In percolation on a lattice, the one dimensional model is trivial. This is
not the case for continuum percolation, where the objects connecting to each 
other may occupy arbitrary positions \cite{bal}. The richness of continuum 
percolation lies in the variety of the binding criterion (which may include 
effects like the shape and spatial distribution of the objects) and the 
existence of interactions. The interplay between the connectivity criterion 
and the interactions make the theory of continuum percolation a very 
challenging field. Therefore, even in one dimension, no completely general 
results can be derived for continuum percolation. However, some specific models
can be solved. The main such model is the one dimensional version of the 
extended spheres system used some time ago to model microemulsions 
\cite{int}.

The system consists of $N+1$ particles on a closed ring of length $L$. 
The particles interact with each other through a pure hard-core potential 
$v(x_i, x_j) \equiv v(i,j)$, \ie, for two particles $i$ and $j$
\be
v(x_i, x_j) = \left\{ \begin{array}{r @{\quad \quad}l}
   \infty & | x_i - x_j | < \si \\   0& | x_i - x_j | > \si \quad ,
   \end{array} \right.   
\label{modelo}
\ee
where $\si$ is the hard core diameter.

The connectivity criterion is supplied by the existence of a soft (also called
permeable) shell of diameter $d$ around each particle. Two particles are bound
if their shells overlap. Letting $p(x_i,x_j) \equiv p(i,j)$ be the probability
that two particles at $x_i$ and $x_j$ are bound, we have
\be
p(x_i, x_j) = \left\{ \begin{array}{r @{\quad \quad}l}
   1 & | x_i - x_j | < d \\   0& | x_i - x_j | > d  \quad .
   \end{array} \right.   
\label{modelt}
\ee
Naturally, $d>\si$. The percolation transition now arises as a function of the 
density $\rho= (N+1)/L$.
The mean cluster size $S$ diverges at a critical density $\rho_c$ which signals
the sharp (in the thermodynamic limit) appearance of an infinite cluster. The
order parameter is $P(\rho)$, the probability that a randomly selected particle
belongs to this infinite cluster.

The solution of this model relies on a general mapping I described recently,
between continuum percolation and a Potts fluid \cite{drory}. The Potts fluid 
is a system of freely moving spins $\{ \la_i\}_{i =1} ^N $ with $s$ states, 
which interact with each other through a spin-dependent potential 
$V(x_i,\la_i;x_j,\la_j)$, such that
\be
V(x_i,\la_i; x_j, \la_j) = \left\{ \begin{array}{r @{\quad \mbox{if} \quad}l}
   U(x_i,x_j) & \la_i =\la_j \\ W(x_i,x_j) & \la_i \neq\la_j   \quad .
   \end{array} \right.  \label {Potts}
\ee
Here $U$ and $W$ are arbitrary functions.

Furthermore, the spins couple to an external field $h(x)$ through an 
interaction hamiltonian
\be
H_{int} = - \sum_{i=1}^N \,\p (\la_i) h(x_i) \quad ,
\ee
where
\be
\p (\la)  =  \left\{ \begin{array}{r @{\quad \mbox{if} \quad} l}
       s - 1 & \la = 1 \\ -1 & \la \ne 1  \quad .\end{array} \right.\label{psi}
\ee

Every continuum percolation model defined by an interaction $v(i,j)$ and a 
binding criterion $p(i,j)$ can be mapped exactly on a Potts fluid \cite{drory}
by choosing

\bea
U(i,j) & = & v(i,j) \quad ,\nonumber\\
\exp \left[- \beta W(i,j) \right] & = & q(i,j) \exp \left[ - \beta v(i,j) 
\right] \quad ,
\label{map}
\eea
where
\be
q(i,j) \equiv 1 - p(i,j)  \quad .
\ee

For a constant field, the Potts configuration integral is
\be
Z = \frac{1}{N!} \sum_{\{\la_m\}} \int \, dx_1 \cdots dx_N \, \exp \left[ 
-\beta \sum_{i>j} V(i,j) + \beta h \sum_{i=1}^N \p (\la_i) \right] \quad ,
\label{conf}
\ee
where the sum $\sum_{\{\la_m\}}$ is performed over all spin configurations, 
and $\beta = 1 / kT$ as usual.

The magnetization of the Potts fluid is defined as 
\be
M = \frac{1}{\beta N (s-1)} \frac{\partial \ln Z}{\partial h} \quad ,
\label{defm}
\ee
and the susceptibility is 
\be
\chi = \frac{\partial M}{\partial h} \quad .
\ee
The Potts 2-density function is defined as
\be
\rho^{(2)} (\x, \mu ; \y, \eta) = \left\langle \sum_{i=1}^N \sum_{{j=1 \atop 
j \ne i}}^N\delta (\x_i -\x) \,\delta (\x_j -\y)\, \delta_{\la_i, \mu}\, 
\delta_{\la_j, \eta} \right\rangle \quad .  \label{defro}
\ee

According to the general mapping between continuum percolation and the Potts 
fluid, the percolation probability and the mean cluster size are obtained as
the limits \cite{drory},
\bea
P(\rho) &=& \lim_{h \to 0} \, \lim_{N \to \infty} \, \lim_{s \to 1} \,
M \quad ,\\
S (\rho) &=& \lim_{h \to 0}\, \lim_{N \to \infty} \, \lim_{s \to 1} \,
\frac{1}{\beta}\chi \quad\qquad (\rho <\rho_c) \quad .
\eea
Hence the strategy is to calculate $M$ and $\chi$ for the adequate Potts fluid
and to obtain from them $P(\rho)$ and $S$ by taking the appropriate limits.
Furthermore, the Potts 2-density function is related to a fundamental quantity
in percolation, the pair-connectedness function, $\gd$, which is defined as
\bea
\rho^2 \,\gd (\x , \y)\, d \x d \y &=& 
\mbox{Probability of finding two particles in regions } \nonumber \\
& & d \x \mbox{ and } d \y \mbox{ around the positions }
\x \mbox{ and } \y \mbox{, such} \nonumber \\
& & \mbox{that they both belong to the same cluster,} \nonumber\\
\eea
where $\rho$ is the numerical density. This function is related to the mean 
cluster size through the relation (for a translationally invariant system)
\be
S = 1 + \frac{1}{N} \int d \x \, d \y \,\,\rho^2 \, \gd (\x, \y) \quad .
\label{Sone}
\ee
The relation to the 2-density function is given by 
\be
\gd (\x, \y) = \lim_{s \to 1} \frac{1}{\rho^2}\left[ \rho^{(2)}( \x,\mu ;
\y, \mu) - \rho^{(2)}(\x,\mu;\y, \eta) \right] \qquad (\rho < \rho_c ) \quad,
\label{gdagf}
\ee
where $\mu, \eta \ne 1$ and $\mu \ne \eta$, but are otherwise arbitrary.

For the percolation model defined be Eqs.~(\ref{modelo}) and (\ref{modelt}), 
the proper Potts fluid is determined by Eq.~(\ref{map}) to be
\bea
\exp\left[ -\beta U(x_i,x_j)\right] \equiv Q(i,j) &=&  
\left\{ \begin{array}
{r @{\quad \quad}l}
   0& | x_i - x_j | < \si \\   1& | x_i - x_j | > \si  
   \end{array} \right.  \quad ,\\
\exp\left[ -\beta W(x_i,x_j)\right] \equiv R(i,j) &=&  \left\{ \begin{array}
{r @{\quad \quad}l}
   0& | x_i - x_j | < d \\   1& | x_i - x_j | > d  \quad .
   \end{array} \right .
  \label{SR}
\eea
What makes this model solvable is that these functions can take only the values
$0$ and $1$ (this condition may be relaxed if the values of $d$ are restricted
to $d<2\si$). Mathematically, this system presents several similarities with 
the Takahashi gas \cite{taka}, and several parts of the following derivation 
follow the calculations of Takahashi and of Salsburg at al. \cite {sal}

\section{Solution of the model}
\setcounter{equation}{0} 

We assume that the spins are ordered on a closed ring of length $L$, so that
the position $0$ and the position $L$ are identified. Along this ring, we place
$N+1$ spins, numbered from $0$ to $N$, so that one of the spins is fixed at 
the position $0$. The property which makes such one dimensional models solvable
is the existence of a canonical ordering of the particles. For a configuration 
$\{ x_0, x_1,\ldots,x_N\}$, the canonical ordering consists in labeling the 
leftmost particle (fixed at the position $0$), as $0$, the one immediately to 
its right as $1$, and so on, until the rightmost particle, labeled $N$. Thus,
\be
0 = x_0 < x_1 < x_2 < \cdots < x_N < L \equiv x_{N+1} \quad .
\ee
The definition $x_{N+1} \equiv L$ serves to simplify the notation later on.
There are exactly $(N+1) !$ configurations which differ from the canonically 
ordered one only by the labels attached to the particles, provided we 
distinguish clockwise numbering from counter-clockwise, \ie, provided the ring
is oriented. Since all the position variables are integrated upon, each of 
these configurations contributes the same to $Z$, which is therefore $(N+1) !$ 
times the contribution of the canonically ordered configuration. Hence,
\bea
Z &=& \!\!\! \int_0^L \!dx_N \!\! \int_0^{x_N} \! d x_{N-1} \cdots 
\int_0^{x_2} dx_1 \sum_{\{\la_m\}} \exp \left[ -\beta \sum_{i>j} V(i,j) + 
\beta h \sum_{i=0}^N \p (\la_i) \right] \nonumber \\
\label{Zone}
\eea
This property is completely general and is independent of the specific form of
$V(i,j)$. Let us denote
\be
f_{\la_i, \la_j} (|x_i - x_j|) \equiv \exp\left[-\beta V(x_i,x_j) \right]= 
\left\{ \begin{array}{r @{\quad \quad}l}
   Q(i,j)& \la_i = \la_j \\    R(i,j)& \la_i \ne \la_j   \quad .
   \end{array} \right. \label{expf}
\ee
Now, taking into account that $x_0 = x_{N+1}$ because of the periodic boundary
conditions, the configuration integral can be written as
\be
Z=  \int_0^L dx_N \int_0^{x_N} d x_{N-1} \!\cdots \!\!
\int_0^{x_2} dx_1 \sum_{\{\la_m\}}\prod_{i>j}^{N} f_{\la_i, \la_j} 
(|x_i - x_j|) \exp\left[ \beta h \sum_{i=0}^N \p (\la_i) \right]\label{ZR}
\ee

The specifics of the model, Eq.~(\ref{SR}), now enter to verify the following 
theorem.

{\bf Theorem (Factorization)}. {\it For any spin configuration } $\{\la_m\}$,
\be
\prod_{i>j=0}^N f_{\la_i, \la_j} (|x_i - x_j|) = \prod_{i=0}^{N-1}
f_{\la_{i+1}, \la_i} (x_{i+1} - x_i) \quad .
\ee

Although simple, this theorem is not entirely trivial because of the
presence of two scales, $\si$ and $d$, the interplay of which depends on the 
spin configuration $\{\la_m\}$. The theorem is proved in the Appendix. Using 
this result, Eq.~(\ref{ZR}) can now be rewritten as
\bea
Z =  \int_0^L \! dx_N \int_0^{x_N}\!\! d x_{N-1} \!\!&\cdots& \!\!\!\!\!
\int_0^{x_2}\! dx_1 \sum_{\{\la_m\}} \Bigg\{f_{\la_0, \la_N} (L-x_N)
\prod_{i=1}^N f_{\la_{i+1}, \la_i} (x_{i+1} - x_i) \nonumber \\
&&\,\,\qquad \qquad \times f_{\la_1, \la_0} (x_1)\exp\left[ \beta h 
\sum_{i=1}^N \p (\la_i) \right] \Bigg\} \quad .
\eea
Let us define
\be
G_{\la_i,\la_j}(y) \equiv f_{\la_i,\la_j}(y) \exp \left\{ \frac{\beta h}{2}
\left[ \p(\la_i) + \p(\la_j) \right] \right\} \quad . \label{dfg}
\ee
Then
\bea
Z =  \int_0^L  dx_N \!\!&\cdots& \!\!\!\! 
\int_0^{x_2} dx_1  \sum_{\{\la_m\}}\Bigg\{G_{\la_0, \la_N} (L-x_N) G_{\la_N,
\la_{N-1}}(x_N - x_{N-1}) \nonumber \\
&& \,\:\quad \qquad \qquad \times \cdots G_{\la_2, \la_1} (x_2 - x_1) 
G_{\la_1, \la_0} (x_1)\Bigg\} \quad . \label{zg}
\eea
This multiple integral has the form of a Laplace convolution. Denoting
\be
T_{\la_i,\la_j}(\om) \equiv \int_0^{\infty} dy \,\, e^{-\om y} \,
G_{\la_i,\la_j}(y) \quad ,      \label{defT}
\ee
we have
\bea
\int_0^{\infty} dL \, e^{-\om L} \,Z(L) &=& \sum_{\{\la_o, \la_1, \ldots, 
\la_N\}} T_{\la_0, \la_N}(\om) T_{\la_N, \la_{N-1}} (\om) \cdots 
T_{\la_2, \la_1} (\om)T_{\la_1, \la_0} (\om) \nonumber \\
&=& \mbox{Tr} \left( T^{N+1} \right) \quad .
\eea
The matrix $T_{\la_i,\la_j}(\om)$ now plays the role of an effective transfer 
matrix for this problem. It is an $s \times s$ matrix, of the form
\be
T = \left( \begin{array}{*{5}{c}}
A & B & B & \cdots & B \\
B & C & D & \cdots & D \\
B & D & C & \cdots & D \\
\vdots & \vdots & \vdots & \ddots & \vdots \\
B & D & D & \cdots & C
\end{array}  \right) \quad ,
\ee
where
\bea
A &=& \int_0^{\infty} dy \, e^{-\om y} Q(y) e^{\beta (s-1) h} = \frac{1}{\om}
\exp\left[ - \om \si + \beta (s-1) h \right] \quad ,\nonumber \\
B &=& \int_0^{\infty} dy \, e^{-\om y} R(y) e^{\frac{1}{2}\beta (s-2) h} = 
\frac{1}{\om}\exp\left[ - \om d + \frac{1}{2}\beta (s-2) h \right] \quad ,
\nonumber\\
C &=& \int_0^{\infty} dy \, e^{-\om y} Q(y) e^{- \beta h} = \frac{1}{\om}
\exp\left[ - \om \si - \beta h \right] \quad , \nonumber \\
D &=& \int_0^{\infty} dy \, e^{-\om y} R(y) e^{-\beta h} = 
\frac{1}{\om}\exp\left[ - \om d - \beta h \right] \quad . \label{ABCD}
\eea
Let $\{ \al_i\}_{i=1}^s$ be the eigenvalues of $T$. Then
\be
\int_0^{\infty} dL \, e^{-\om L} Z(L) = \sum_{i=1}^s \, \al_i^{N+1} \quad .
\label{LTZ}
\ee
The eigenvalues are found easily to be
\bea
\al_1 &=& \frac{1}{2}\left[ A + C + (s-2)D + \sqrt{\Delta} \right] \quad ,
\nonumber \\
\al_2 &=& \frac{1}{2}\left[ A + C + (s-2)D - \sqrt{\Delta} \right] \quad ,
\nonumber \\
\al_3 &=& \cdots = \al_s = C-D \quad ,
\label{eigen}
\eea
where
\be
\Delta = \left[ A - C - (s-2)D \right]^2 + 4 (s-1) B^2 \quad .
\ee

Reversing the Laplace transform in Eq.~(\ref{LTZ}) yields
\be
Z = \sum_{j=1}^s \, \frac{1}{2 \pi i} \int_{a - i \infty}^{ a + i \infty} d \om
\, \exp \left[ \om L + (N+1)\ln \al_j (\om) \right] \quad .\label{Zrev}
\ee
In the thermodynamic limit, the integral can be evaluated {\it exactly} by 
steepest descent. Let us introduce the notation
\be
\g_j(\om) \equiv \om \frac{L}{N+1} + \ln \al_j (\om) \quad .\label{defg}
\ee
Then the maximum of the exponents in Eq.~(\ref{Zrev}) is obtained from the
condition
\be
\frac {d \g_j}{d \om} \Bigg\vert_{\om = \om_j} = 0 \quad , \label{defom}
\ee
which defines the quantity $\om_j$ at which $\g_j (\om)$ is maximal. From the 
usual theory of steepest descent \cite{orszag}, we now obtain from
Eq.~(\ref{Zrev}) that
\be
Z = \sum_{j=1}^s K_j \exp\left[(N+1) \g_j(\om_j)\right] \quad , \label{Zfin}
\ee
where 
\bea
K_j &=& \frac{1}{\left[2 \pi (N+1) \g_j^{''} (\om_j)\right]^{1/2}} \quad ,
\nonumber \\
\g_j^{''} (\om) &=& \frac{d^2 \g_j (\om)}{d \om ^2} \quad .\label{kj}
\eea

The sum in Eq.~(\ref{Zfin}) is dominated by $\g_1$. To see this, note first 
that since $\Delta > 0$, we have $\al_1 > \al_2$. Furthermore,
\be
\al_1 - \al_3 = \frac{1}{2} \left[ A - C + s D + \sqrt{\Delta} \right] \quad.
\ee
Since $ A = C e^{ s \beta h}$ and $ h> 0$, we have that $A-C > 0$. Therefore, 
$\al_1 - \al_3> 0$. Hence, for any given $\om$, $\al_1 (\om) > \al_j (\om)$ 
for $j = 2, \dots, s$. Since $ \, \g_1(\om) - \g_j (\om) = \ln \left( \al_1 /
\al_j\right)$, we also have that for any given $\om$, $\, \g_1 (\om) > \g_j 
(\om)$ for $ j = 2, \dots, s$. Now, at $\om_1$, $\g_1$ is maximal, so that
\be
\g_1(\om_1) \ge \g_1 (\om_j) > \g_j (\om_j)
\ee
for $ j = 2, \dots, s$. Hence, for $N \to \infty$
\be
Z = K_1 e^{N \g_1} \left\{ 1 + \sum_{j=2}^s \frac{K_j}{K_1} \exp\left[N \g_1 
(\om_1) - N \g_j(\om_j)\right]\right\} \sim  K_1 e^{N \g_1} \quad .
\ee
Hence,
\be
\lim_{N \to \infty} \frac{1}{N} \ln Z = \g_1 \quad . \label{Zs}
\ee

From Eq.~(\ref{Zs}), we can obtain the magnetization $M$,
\bea
M = \frac{1}{\beta (s-1) N} \frac{ d \ln Z}{ d h} &=& \frac{1}{\beta (s-1) N}
\left[ \left( \frac{\partial \g_1}{\partial h} \right)_{\om_1} + 
\left( \frac{\partial \g_1}{\partial \om_1} \right)_{h}\left( \frac{\partial 
\om_1}{\partial h} \right) \right]  \nonumber \\
&=& \frac{1}{\beta (s-1)} \left( \frac{\partial \g_1}{\partial h} 
\right)_{\om_1} \quad , \label{mags}
\eea
where we made use of the fact that $0 = \left( \partial \g_1 / \partial \om_1 
\right)_{h} $ by definition of $\om_1$.

Although Eq.~(\ref{Zs}) is in principle the exact solution of the one 
dimensional Potts fluid, it is not explicit enough to be useful. However, we
are not interested in the Potts fluid itself, but rather in the percolation 
model. This is obtained in the limits $ s \to 1$ and $ h \to 0$, and in these
limits all quantities can be calculated explicitly.

To do this, we set $ s = 1+ \ep$, and calculate all relevant quantities to
first order only in $\ep$ (this turns out to be sufficient). Therefore, from
Eqs.~(\ref{ABCD}) and (\ref{defg}), we find after some algebra that
\be
\g_1 = \om \frac{L}{N} + \ln A + \ep \left[ \beta h + \frac{B^2}{A(A-C+D)}
\right] + O( \ep ^2) \quad ,
\label{gep}
\ee
where $A,\,B,\,C,\,D\,$ must be evaluated to zero order in $\ep$. From 
Eq.~(\ref{ABCD}), the result for this is
\bea
A &=& \frac{1}{\om} \exp\left[ - \om \si \right] \quad , \nonumber \\
B &=& \frac{1}{\om} \exp\left[ - \om d - \frac{1}{2} \beta h \right]  \quad ,
\nonumber \\
C &=& \frac{1}{\om} \exp\left[ - \om \si - \beta  h \right] \quad , 
\nonumber \\
D &=& \frac{1}{\om} \exp\left[ - \om d - \beta h \right] \quad . \label{abcd}
\eea

Now,
\be
M = \lim_{\ep \to 0} \frac{1}{\beta \ep} \left( \frac{\partial \g_1}
{\partial h} \right)_{\om_1} = 1 + \frac{1}{\beta} \frac{\partial{ }}
{\partial h} \left[ \frac{B^2}{A(A-C + D)} \right]_{\om_1, \ep =0} \quad.
\ee
After some algebra, and with the use of Eq.~(\ref{abcd}), we end up with
\be
M = 1 - \left[ \frac{B}{A-C + D} \right]_{\om_1, \ep =0}^2 \quad .\label{mperc}
\ee
After some more algebra, the susceptibility is obtained straightforwardly,
\be
\chi = \frac{\partial M}{\partial h} = \beta \left[\frac{B^2(A+C-D)}
{(A-C + D)^2} \right]_{\om_1, \ep=0} \quad . \label{chiperc}
\ee
In both Eq.~(\ref{mperc}) and Eq.~(\ref{chiperc}), $\om_1$ is evaluated at 
$\ep = 0$. From Eqs.~(\ref{gep}) and (\ref{abcd}), we have that for $\ep = 0$,
\be
\g_1 = \frac{\om}{\rho} - \om \si -\ln \om \quad ,
\ee
where $\rho = (N+1)/L$ is the density. Hence from $\left( d \g_1 / d \om 
\right)_{\om = \om_1} = 0 $, we obtain
\be
\om_1 \Bigg\vert_{\ep = 0} = \frac{\rho}{1 - \rho \si} \quad .\label{om1}
\ee

The final result is obtained from Eqs.~(\ref{mperc}) and (\ref{chiperc}) by 
setting $\ep = 0$, plugging in the value of $\om_1$ and taking the limit $ h\to
0$. We have
\bea
M \to P(\rho) &=& 1 - \lim_{h \to 0}\left\{1 - \left(e^{\beta h} - 1 \right)
\exp\left[\frac{\rho(d - \si)}{1 - \rho \si} \right]\right\}^{-2} \quad ,
\label{pf}\\
\frac{\chi}{\beta} \to S(\rho) &=& 2\exp\left[\frac{\rho(d - \si)}
{1 - \rho \si} \right] - 1  \quad .\label{sf}
\eea

If we just set $h = 0$ in the equation for $P(\rho)$, we obtain $P(\rho) = 0$,
which shows there is no transition. There is however, one exception, obtained
at $\rho = 1/\si = \rho_{cp}$. This is the close packing density in one 
dimension, and clearly we must have a percolating cluster in this case. Indeed,
at this density
\be
\lim_{\rho \to \rho_{cp}} = \exp\left[\frac{\rho(d - \si)}{1 - \rho \si} 
\right] = \infty \quad ,
\ee
and therefore
\be
\left\{1 - \left(e^{\beta h} - 1 \right) \exp\left[\frac{\rho(d - \si)}
{1 - \rho \si} \right]\right\}^{-2} \Bigg\vert_{\rho = \rho_{cp}} = 0
\ee
for {\it all} values of $h$; hence, $P(\rho_{cp}) = 1 $ as expected. Note that 
for $\si = 0$ (no hard core), $\rho_{cp} \to \infty$. Eq.~(\ref{pf}) becomes in
this case
\be
P(\rho) = 1 - \lim_{h \to 0}\left[1 - \left(e^{\beta h} - 1 \right) e^{\rho d}
\right]^{-2} \quad ,
\ee
and indeed $P(\rho = \infty) = 1$.

The transition is clearer in the expression Eq.~(\ref{sf}) for the mean cluster
size.
\be
\lim_{\rho \to \rho_{cp}} S(\rho) = \lim_{\rho \to \rho_{cp}} 2\exp
\left[\frac{\rho(d - \si)} {1 - \rho \si} \right] - 1  = \infty \quad,
\ee
so that indeed the mean cluster size diverges as the density approaches the 
close packing value.

\section{The pair-connectedness function}
\setcounter{equation}{0} 

Let us now consider the pair connectedness. To this end we need to calculate 
the spin 2-density function, 
\be
\rho^{(2)} (x, \mu; y, \eta) = \left\langle \sum_{i=1}^N \sum_{{j=1 
\atop j \ne i}}^N\delta (x_i - x) \,\delta ( x_j - y)\,\delta_{\la_i, \mu}
\, \delta_{\la_j, \eta} \right\rangle \quad .  \label{defro2}
\ee
To simplify the algebra let us use the system's translational invariance (one 
can work with Eq.~(\ref{defro2}) and obtain translational invariance directly,
but this adds needlessly to the calculations). We assume that $y < x$, denote 
$x - y = r$ and set $ y=0$ and $\la_0 = \mu$. We now have, therefore, that
\be
\rho^{(2)} (0, \mu; r, \eta) \equiv \rho^{(2)} (r ; \mu, \eta) = \rho
\left\langle \sum_{k=1}^N \delta (x_k - r) \,\delta_{\la_0, \mu}\, 
\delta_{\la_k, \eta} \right\rangle \quad , \label{defro3}
\ee
where $\rho = (N+1)/L$. From Eq.~(\ref{zg}), we have that
\bea
\rho^{(2)} (r ; \mu, \eta) &=& \frac{\rho}{Z}\sum_{k=1}^N \, \int_0^L  dx_N 
\cdots \int_0^{x_2} dx_1  \nonumber \\
&\times& \sum_{\{\la_m\}}\,\Bigg\{\delta_{\la_0, \mu} \, G_{\la_0, \la_N} 
(L-x_N) G_{\la_N,\la_{N-1}}(x_N - x_{N-1}) \nonumber \\
&\times & \cdots G_{\la_{k+1},\la_k}(x_{k+1} - x_k)
\delta_{\la_k, \eta} \delta (x_k - r) G_{\la_k,\la_{k-1}}(x_k - x_{k-1}) 
\nonumber \\
&\times& \cdots G_{\la_2, \la_1} (x_2 - x_1) G_{\la_1, \la_0} (x_1)\Bigg\}
\quad ,
\eea
where the function $G_{\la_i, \la_j}(x)$ is defined in Eq.~(\ref{dfg}). This  
is again a Laplace convolution and we have
\bea
\int_0^{\infty} dL \, e^{-\om L} \,\frac{Z}{\rho} \, \rho^{(2)}(r ; \mu, \eta)
= \sum_{k=1}^N \sum_{\{\la_0, \la_1, \ldots, \la_N\}} &&\!\!\!\!\!\!\!\!\!\!\!
\delta_{\la_0, \mu}T_{\la_0, \la_N}(\om) \cdots T_{\la_{k+1}, \la_k} (\om) 
\nonumber \\
&\times& \delta_{\la_k, \eta} \, e^{-\om r} \, h_k (r; \la_k , \la_0) \quad ,
\nonumber
\eea
where $h_k (r; \la_k , \la_0)$ is defined by its Laplace transform as
\be
\int_0^{\infty} dr \, e^{-\nu r} \,h_k(r ; \la_k, \la_0) = \sum_{\{\la_{k-1},
\ldots, \la_1\}} T_{\la_k, \la_{k-1}}(\nu) \cdots T_{\la_1, \la_0} (\nu) = 
\left( T^{k} \right)_{\la_k, \la_0} \quad . \label{defh}
\ee
The sum over the spins yields
\be
\int_0^{\infty} dL \,\, e^{-\om L} \,\frac{Z}{\rho} \rho^{(2)}(r ; \mu, \eta)
= \sum_{k=1}^N \left( T^{N-k +1}(\om) \right)_{\mu, \eta} e^{-\om r} \,
h_k (r; \eta , \mu) \quad .\label{four}
\ee
Inverting the Laplace transform in Eqs.~(\ref{defh}) and (\ref{four}), we 
obtain finally
\bea
\frac{Z}{\rho} \rho^{(2)}(r ; \mu, \eta) = \frac{1}{2 \pi i} 
\int_{a - i \infty}^{a + i \infty} d \om \, \frac{1}{2 \pi i} \int_{a - i 
\infty}^{a + i \infty} &d \nu& \!\!\! \sum_{k=1}^N \left[ 
T^{N-k +1}(\om)\right]_{\mu, \eta} \nonumber \\
&\times& \left[ T^{k}(\nu) \right]_{\eta , \mu} e^{\om (L -r)}e^{\nu r} \quad .
\nonumber \\ \label{lptr}
\eea

Let us introduce now a basis of eigenvectors of the matrix $T(\zeta)$, which 
will be denoted $\left\{ \langle u_1 \vert ,\langle u_2 \vert, \ldots, 
\langle u_s \vert \right\}$, $\langle u_j \vert$ being associated with the 
eigenvalue $\al_j$ defined in Eq.~(\ref{eigen}). Straightforward calculation 
yields that
\bea
\langle u_1 \vert &=& \frac{\left( B(s-1), \al_1 - A, \al_1 - A, \ldots, 
\al_1 - A\right)}{\left[B^2(s-1)^2 + (s-1)(\al_1 - A)^2 \right]^{1/2}} \quad,
\nonumber \\
\langle u_2 \vert &=& \frac{\left( B(s-1), \al_2 - A, \al_2 - A, \ldots, \al_2 
- A\right)} {\left[B^2(s-1)^2 + (s-1)(\al_2 - A)^2 \right]^{1/2}} \quad,
\nonumber \\
\langle u_3 \vert &=& \frac{1}{\sqrt{2}}\left(0, 1, -1, 0, \ldots, 0\right) 
\quad , \nonumber \\
\langle u_4 \vert &=& \frac{1}{\sqrt{6}}\left(0, 1, 1, -1, 0, \ldots, 0\right) 
\quad, \nonumber \\
\vdots \nonumber \\
\langle u_s \vert &=& \frac{1}{\sqrt{(s-1)(s-2)}}\left(0, 1, 1,\ldots, 
-(s-2)\right) \quad ,
\eea
where $A, B, \al_1, \al_2$ are defined in Eqs.~(\ref{ABCD}) and (\ref{eigen}).
These expressions can be simplified, however, by setting the magnetic field 
$h = 0$, since it plays no part in the calculation of the pair connectedness. 
With this, we have that $A=C$ and $B=D$. From Eqs.~(\ref{ABCD}) and 
(\ref{eigen}), we now have that
\bea
\al_1 &=& A + (s-1)B \quad ,\nonumber \\
\al_2 &=& A - B \quad , \nonumber \\
A(\zeta) &=& \frac{1}{\zeta} \exp ( - \zeta \si) \quad , \nonumber \\
B(\zeta) &=& \frac{1}{\zeta} \exp ( - \zeta d)   \quad . \label{zeroh}
\eea
Hence,
\bea
\langle u_1 \vert &=& \frac{1}{\sqrt{s}}\left(1, 1, 1, \ldots, 1\right) 
\quad , \nonumber \\
\langle u_2 \vert &=& \frac{1}{\sqrt{s(s-1)}}\left(s-1, -1, -1, \ldots, -1 
\right) \quad ,\nonumber \\
\langle u_3 \vert &=& \frac{1}{\sqrt{2}}\left(0, 1, -1, 0, \ldots, 0\right) 
\quad , \nonumber \\
\vdots \nonumber \\
\langle u_s \vert &=& \frac{1}{\sqrt{(s-1)(s-2)}}\left(0, 1, 1,\ldots, 
-(s-2)\right) \quad .
\label{eigenv}
\eea
Finally, we define two vectors of the type $(0,0,\ldots,1,0,\ldots,0)$, denoted
$\langle \mu \vert $ and $\langle \eta \vert$ such that their single non zero 
component `` 1 " stands at the position $\mu$ and $\eta$ respectively.
With this, we have that
\bea
\left[ T^{N-k +1}(\om) \right]_{\mu, \eta} &=& \sum_{m=1}^s \langle \mu \vert 
u_m \rangle \, \al_m^{N-k+1} (\om) \, \langle u_m \vert \eta \rangle \quad ,
\nonumber \\
\left[ T^k(\nu) \right]_{\eta, \mu} &=& \sum_{n=1}^s \langle \eta \vert 
u_n \rangle \, \al_n^k (\nu) \, \langle u_n  \vert \mu \rangle  \quad .
\label{power}
\eea
Hence,
\bea
\lefteqn{\frac{1}{2 \pi i} \int_{a - i \infty}^{a + i \infty} \, d \om \left[ 
T^{N-k +1}(\om)\right]_{\mu, \eta}e^{\om (L -r)}} \nonumber \\
&=& \frac{1}{2 \pi i} 
\int_{a - i \infty}^{a + i \infty}  d \om \,  e^{\om L} \sum_{m=1}^s \, 
\al_m^{N+1} (\om)\, F_{m,k} (\om, r) \nonumber \\
&=& \sum_{m=1}^s \frac{1}{2 \pi i} \int_{a - i \infty}^{a + i \infty} 
d \om \, e^{(N+1)\g_m(\om)} F_{m,k} (\om, r) \quad,
\eea
where
\be
F_{m,k} (\om,r) = \frac{\langle \mu \vert u_m \rangle \langle u_m \vert \eta 
\rangle}{\left[ \al_m (\om)\right]^k} e^{ - \om r} \quad ,
\ee
and $\g_m(\om) = \om L /(N+1) + \ln \al_m (\om)$ as in Eq.~(\ref{defg}). In the
limit $N, L \to \infty$ the integral can be evaluated exactly by steepest
descent. The calculation is identical to the one performed in 
Eqs.~(\ref{Zrev})-(\ref{Zfin}). The result is
\bea
\lefteqn{\lim _{N \to \infty} \frac{1}{2 \pi i} \int_{a - i \infty}^{a + i 
\infty} d \om \left[ T^{N-k +1}(\om)\right]_{\mu, \eta}e^{\om (L -r)}}
\nonumber \\ 
&=& \lim_{N \to \infty} \sum_{m=1}^s K_m \exp\big[(N+1) \g_m(\om_m)\big] 
F_{m,k} (\om_m, r) \quad ,
\eea
with $K_m$ and $\om_m$ defined in Eqs.~(\ref{kj}) and (\ref{defom}). As in 
Eq.~(\ref{Zs}), the term $\g_1$ dominates all the others, so that in the limit
$N \to \infty$, we have, exactly,
\bea
\lefteqn{\lim _{N \to \infty} \frac{1}{2 \pi i} \int_{a - i \infty}^{a + i 
\infty}  d \om \left[ T^{N-k +1}(\om)\right]_{\mu, \eta}e^{\om (L -r)}}
\nonumber \\
&=& \lim_{N \to\infty} K_1 \exp\left[(N+1) \g_1(\om_1)\right] F_{1,k} 
(\om_1, r) \quad . \label{star}
\eea
From Eq.~(\ref{Zs}), we have that, in the same limit, $Z = K_1 \exp [ (N+1)
\g_1 ( \om_1)] $. Therefore, combining Eqs.~(\ref{lptr}), (\ref{power}) and
(\ref{star}), we finally obtain the expression
\bea
&&\rho^{(2)}(r ; \mu, \eta) = \lim_{N \to \infty} \,  \rho \sum_{k=1}^N 
F_{1,k} (\om_1,r) \frac{1}{2 \pi i} \int_{a - i \infty}^{a + i \infty} 
d \nu  \left[ T^{k}(\nu) \right]_{\eta , \mu} e^{\nu r} \nonumber \\
&& \qquad=\rho \sum_{k=1}^{\infty}\frac{\langle \mu \vert u_1 \rangle 
\langle u_1 \vert \eta \rangle}{2 \pi i} \int_{a - i \infty}^{a + i \infty}
d \nu \sum_{n=1}^s \langle \eta \vert u_n \rangle \langle u_n  \vert \mu 
\rangle \left[ \frac{\al_n (\nu)}{\al_1 (\om_1)}\right]^k \, e^{(\nu - \om_1)r}
\quad . \nonumber \\
\label{ro2}
\eea

In order to calculate the pair connectedness, we need the case $\mu, \eta, \ne
1$ [see Eq.~(\ref{gdagf})]. For this case, we see directly from the definition
of the $\langle u_i \vert$, Eq.~(\ref{eigenv}), that for any $ \la \ge 2$,
\bea
\langle u_1 \vert \la \rangle &=& \frac{1}{\sqrt{s}} \quad ,\nonumber \\
\langle u_2 \vert \la \rangle &=& \frac{-1}{\sqrt{s(s-1)}} \quad ,\label{prod1}
\eea
and for any $n \ge 3$, 
\be
\langle u_n \vert \la \rangle = \left\{ \begin{array}{r @{\quad \mbox{if} 
\quad}l}
  0 & n <\la \\ 
 \frac{\displaystyle {-(n-2)}}{\displaystyle{\sqrt{(n-2)(n-1)}}}  &  n = \la\\ 
  \frac{\displaystyle {1}}{\displaystyle{\sqrt{(n-2)(n-1)}}} &  n > \la  
   \quad . \\
   \end{array} \right.   \label {prod2}
\ee
Hence,
\be
\rho^{(2)}(r ; \mu, \eta) = \frac{\rho}{s} \sum_{k=1}^{\infty}\frac{1}{2 \pi i}
\int_{a - i \infty}^{a + i \infty} d \nu
\sum_{n=1}^s \langle \eta \vert u_n \rangle \langle u_n  \vert \mu \rangle
\left[ \frac{\al_n (\nu)}{\al_1 (\om_1)}\right]^k \, e^{(\nu - \om_1)r} \quad .
\ee
We are interested, however, in the pair connectedness,which involves the 
following difference,
\bea
\lefteqn{\rho^{(2)}(r ; \mu, \mu) - \rho^{(2)}(r ; \mu, \eta)} \nonumber \\
 &=& \frac{\rho}{s} \sum_{k=1}^{\infty}\frac{1}{2 \pi i}\int_{a - i 
\infty}^{a + i \infty}  d \nu \sum_{n=1}^s \left[ \langle \mu \vert u_n 
\rangle - \langle \eta \vert u_n \rangle \right] \langle u_n  \vert \mu 
\rangle \left[ \frac{\al_n (\nu)}{\al_1 (\om_1)}\right]^k \, e^{(\nu - \om_1)r}
\, . \nonumber \\
\eea
From Eq.~(\ref{prod1}), we see that $\langle \mu \vert u_n \rangle - \langle 
\eta \vert u_n \rangle $ vanishes identically for $n = 1, 2 $. Now, for $n \ge
3$, we have that
\be
\al_n (\nu) = \frac{1}{\nu} \left[ e^{-\nu \si} - e^{-\nu d} \right] \quad ,
\ee
and is independent of the value of $n$. Finally, from Eq.~(\ref{prod2}), it is
easily seen that for any $\mu, \eta \ne 1$, $\mu \ne \eta$, we have that
\be
\sum_{n=3}^s  \langle u_n  \vert \mu \rangle  \left[ \langle \mu \vert u_n 
\rangle - \langle \eta \vert u_n \rangle \right] = 1 \quad .
\ee
Therefore, we have that
\be
\rho^{(2)}(r ; \mu, \mu) - \rho^{(2)}(r ; \mu, \eta)= \frac{\rho}{s} 
\sum_{k=1}^{\infty}\frac{1}{2 \pi i}\int_{a - i \infty}^{a + i \infty}
d \nu \left[ \frac{\al_n (\nu)}{\al_1 (\om_1)}\right]^k \, e^{(\nu - \om_1)r}
\label{twoth}
\ee
Finally, to obtain $\gd (r)$, we take the limit $s \to 1$. In this limit,
Eq.~(\ref{zeroh}) yields
\be
\al_1 (\om_1) = \frac{1}{\om_1} \exp \left[ -\om_1 \si \right]  \quad,
\ee
where, from Eq.~(\ref{om1}),
\be
\om_1 = \frac{\rho}{1 - \rho \si} \quad.
\ee
Hence, substituting the values of $\al_n$ and $\al_1$, we finally obtain , by
substituting Eq.~(\ref{twoth}) into Eq.~(\ref{gdagf}), that
\bea
\gd(r) &=& \lim_{s \to 1} \frac{1}{\rho^2}\left[ \rho^{(2)}(r ; \mu, \mu) - 
\rho^{(2)}(r ; \mu, \eta) \right] \nonumber \\
&=& \frac{1}{\rho} \sum_{k=1}^{\infty} \frac{1}{2 \pi i}\int_{a - i 
\infty}^{a + i \infty} d \nu \, \, \frac{\om_1^k}{\nu^k} \, \left[ 1 - 
e^{-\nu (d - \si)} \right]^k   e^{(\nu - \om_1)r} \quad . \label{glp}
\eea
or
\be
\gd(r) = \frac{1}{\rho} \sum_{k=1}^{\infty}\sum_{j=0}^k \, (-1)^j {k \choose j}
\om_1^k \, e^{-\om_1 ( r - k \si)} \,\frac{1}{2 \pi i}\int_{a - i \infty}^{a 
+ i \infty} d \nu \, \frac{1}{\nu^k} \,e^{\nu [r - k \si - j (d- \si)]}
\ee
This integral is an immediate inverse Laplace transform. The final result is
therefore
\bea
\gd(r)= \frac{1}{\rho} \sum_{k=1}^{\infty}\sum_{j=0}^k && \!\!\!\!\!\!\! 
(-1)^j {k \choose j}\left(\frac{\rho}{1 - \rho \si}\right)^k  \, 
\frac{\left[r - k \si - j(d- \si)\right]^{k-1}}{(k-1) !} \nonumber \\
&\times& \Theta [r - k \si + j (d- \si)] \, \exp\left[- 
\frac{\rho ( r - k \si)}{1 - \rho \si} \right]  \label{gfinal}
\eea
where $\Theta (z)$ is the step function
\be
\Theta (z)= \left\{ \begin{array}{r @{\quad \quad}l}
   1 & z > 0 \\ 0 &  z < 0\\ 
   \end{array} \right. \quad .
\ee
We can understand the meaning of $k$ and $j$ as follows. Given a distance 
$r>0$ from the (arbitrary) origin, $\gd (r)$ is a sum of contributions from 
sets of configurations indexed by $k$. The derivation shows that in each such 
configuration there are exactly $k$ particles within the interval $r$. Since 
the potential includes a hard core of diameter $\si$, we expect a condition on
$k$ such that $r > k \si$. This is indeed contained in the expression
$ \Theta [r - k \si + j (d- \si)]$, since it implies that $r > k\si+j(d - \si)
> k \si$. Next, we note that each set of configurations containing $k$ 
particles in the interval $r$ is further divided into subsets indexed by $j$.
The meaning of $j$ is made clear from the condition $ r > k\si+j(d - \si) = 
j d + (k - j) \si $ generated by the step function. This implies that $j$ 
particles out of the $k$ are singlets, \ie, are { \it not} connected to any
other particle in the set. Therefore the step function shows that $\gd (r)$ is
a sum of separate contributions from all sets of configurations containing $k$ 
particles in the interval $r$, $j$ of which are not connected to any other 
particle in the set.

Finally, we can check the expected relation between the mean cluster size $S$
and the pair connectedness. This is worked out  most conveniently from 
Eq.~(\ref{glp}). From this, we have that
\bea
\int_0^{\infty} d r \,\gd (r) &=& \frac{1}{\rho} \sum_{k=1}^{\infty}
\left\{ \frac{\om_1^k}{\nu^k} \, \left[ 1 - e^{-\nu (d - \si)} e^{(\nu - \om_1)
r}\right]^k \right\}_{\nu = \om_1} \nonumber \\
&=& \frac{1}{\rho}  \exp\left[ \frac{\rho (d -\si)}{1 - \rho \si} \right] - 1
\quad .
\eea
From Eq.~(\ref{Sone}), and remembering that we have calculated $\gd (x,y)$
under the assumption that $y < x$, we now have that
\be
S = 1 + \frac{1}{N} \int d x \, d y \,\,\rho ^2 \, 
\gd (x, y) = 1 + 2 \rho \int_0^{\infty} dr \, \gd (r) \quad.
\ee
and therefore
\be
S = 2 \exp\left[\frac{\rho (d -\si)}{1 - \rho \si} \right] - 1 \quad.
\ee
in agreement with the previously obtained result, Eq.~(\ref{sf}).

\section{Summary}
\setcounter{equation}{0} 

I have obtained the exact solution of a one-dimensional percolation model which
includes the effects of interactions among the particles, modeled by a hard 
core  of diameter $\si$. The particles are connected to each other if their 
centers are closer than a distance $d$. The clustering depends on the density 
$\rho$ of the particles. By mapping this system on a Potts fluid, one can 
calculate the mean cluster size, pair connectedness and percolation probability
of this system. As with other one dimensional systems, the phase transition
occurs only in some extreme circumstance, in this case, only when the density
reaches the close-packing value $\rho_{cp} = 1/\si$. As the density increases
towards this critical value, the mean cluster size increases as
\be
S = 2 \exp\left[\frac{\rho (d -\si)}{1 - \rho \si} \right] - 1 \quad ,
\ee
and diverges at $\rho_{cp}$. Meanwhile, the pair connectedness is given by
\bea
\gd(r)= \frac{1}{\rho} \sum_{k=1}^{\infty}\, \sum_{j=0}^k && \!\!\!\!\! \!\!\!
(-1)^j {k \choose j}\left(\frac{\rho}{1 - \rho \si}\right)^k  \, 
\frac{\left[r - k \si - j(d- \si)\right]^{k-1}}{(k-1) !} \nonumber \\
&\times& \Theta [r - k \si + j (d- \si)] \, \exp\left[- 
\frac{\rho ( r - k \si)}{1 - \rho \si} \right]  \quad .
\eea

These results should be particularly useful as a test of approximation methods
or numerical calculations or simulations. Such methods can be checked against
the exact results and some idea thus obtained of their reliability and 
effectiveness. Once a method has proved to be successful in the one dimensional
case, it can be applied to higher dimensions with some hope of success.

From a more general point of view, the present results are a confirmation of
the power provided by the mapping between continuum percolation and the Potts
fluid. Recently, Cinlar and Torquato have used renewal theory to investigate
one dimensional continuum percolation, but without interactions (\ie, the 
$\si \to 0$ limit of the present model) \cite{cinlar}. There seems to be no
simple way of extending this type of direct probability arguments to cover
interactions as well. It appears that in order to discuss general models, 
which include inter-particle interactions, there is at the moment no method 
more powerful than the mapping with the Potts fluid.

\section{Appendix}
\setcounter{equation}{0}
\renewcommand{\theequation}{A.\arabic{equation}}

The proof of the factorization theorem is obtained by induction on $N$. For 
$N = 1$ the theorem is obvious. Let's assume it is correct for $N$ and prove it
for $N+1$. Then
\bea
\lefteqn{\prod_{i>j = 0}^{N+1} f_{\la_i,\la_j} (|x_i - x_j|) } \nonumber \\
&=& \prod_{i>j = 0}^N 
f_{\la_i,\la_j} (|x_i - x_j|)\prod_{i= 0}^N f_{\la_{N+1},\la_i} (x_{N+1}-x_i)
\nonumber\\
&=&\prod_{i= 0}^{N-1} f_{\la_{i+1},\la_i} (x_{i+1}-x_i)\prod_{i= 0}^N 
f_{\la_{N+1},\la_i} (x_{N+1}-x_i) \nonumber \\
&=&\prod_{i=0}^{N-1} \left[f_{\la_{i+1},\la_i} (x_{i+1}-x_i)f_{\la_{N+1},\la_i}
(x_{N+1}-x_i)\right]f_{\la_{N+1},\la_N} (x_{N+1}-x_N) \, , \label{start} 
\nonumber\\
\eea
where we used the induction assumption to obtain the second equality.

Let us now consider the various possibilities for every term $i$.
\begin{enumerate}
\item 
$
f_{\la_{N+1},\la_i}(x_{N+1}-x_i) = 1 \, .
$

In this case clearly,
\be
f_{\la_{i+1},\la_i} (x_{i+1}-x_i)f_{\la_{N+1},\la_i} (x_{N+1}-x_i) = 
f_{\la_{i+1},\la_i} (x_{i+1}-x_i)   \quad . \label{cono}
\ee
\item
$
f_{\la_{N+1},\la_i}(x_{N+1}-x_i) = 0 \mbox{\quad\, and \quad \,} 
\la_{N+1} = \la_i \, .
$

Then, from Eq.~(\ref{expf}),
\be
f_{\la_{N+1},\la_i}(x_{N+1}-x_i) = Q(x_{N+1}-x_i) = 0 \quad,
\ee
so that, necessarily [see Eq.~(\ref{modelo}], $x_{N+1} -x_i < \si$. However, 
in the canonical ordering, $x_i<x_{i+1}<\cdots<x_N$, so that
\be
x_{N+1} - x_N < x_{N+1} - x_i < \si < d \quad .
\ee
Hence,
\be
f_{\la_{N+1},\la_N}(x_{N+1}-x_N) = 0 \quad ,
\ee
even if $\la_{N+1} \ne \la_N$, because $\si < d$. Therefore,
\bea
0 &=& f_{\la_{i+1},\la_i} (x_{i+1}-x_i)f_{\la_{N+1},\la_i} (x_{N+1}-x_i)
f_{\la_{N+1},\la_N}(x_{N+1}-x_N) \nonumber \\
&=& f_{\la_{i+1},\la_i} (x_{i+1}-x_i)f_{\la_{N+1},\la_N}(x_{N+1}-x_N)  \quad.
\label{cont}
\eea
\item
$
f_{\la_{N+1},\la_i}(x_{N+1}-x_i) = 0 \mbox{\,\quad and \,\quad} 
\la_{N+1} \ne \la_i \, .
$

Then [see Eq.~(\ref{expf}],
\be
f_{\la_{N+1},\la_i}(x_{N+1}-x_i) = R(x_{N+1}-x_i) = 0 \quad.
\ee
and, necessarily [see Eq.~(\ref{modelo}], $x_{N+1} -x_i < d$. Then there must 
exist some $j$, $i \le j < N+1$, such that
\be
\la_{N+1} = \la_k \quad \mbox{and} \quad \la_{N+1} \ne \la_j\quad 
\mbox{for all} \quad j < k <+ N+1 \quad ,
\ee
\ie, $j$ is the closest spin to $\la_{N+1}$ which differs from it. In 
particular, $\la_{N+1} = \la_{j+1}$, so that
\be
f_{\la_{j+1},\la_j}(x_{j+1}-x_j) = R(x_{j+1}-x_j) \quad.
\ee
Because of the canonical ordering,
\be
x_{j+1} - x_j \le x_{j+1} - x_i < x_{N+1} - x_i  < d \quad .
\ee
Hence
\be
 R(x_{j+1}-x_j) = 0 \quad .
\ee
Therefore,
\bea
0 &=& f_{\la_{i+1},\la_i} (x_{i+1}-x_i)f_{\la_{N+1},\la_i} (x_{N+1}-x_i)
f_{\la_{j+1},\la_j}(x_{j+1}-x_j) \nonumber \\
&=& f_{\la_{i+1},\la_i} (x_{i+1}-x_i)f_{\la_{j+1},\la_j}(x_{j+1}-x_j)   
\quad .\label{conth}
\eea
\end{enumerate}

From Eqs.~(\ref{cono})-(\ref{conth}), it follows that $f_{\la_{N+1},\la_i} 
(x_{N+1}-x_i)$ makes no difference anywhere in the product $\prod_{i= 1}^{N-1} 
f_{\la_{i+1},\la_i} (x_{i+1}-x_i)f_{\la_{N+1},\la_i} (x_{N+1}-x_i)$. It can 
therefore be dropped out of every term $i$. Hence,
\bea
\lefteqn{\prod_{i= 1}^{N-1} \left[f_{\la_{i+1},\la_i} (x_{i+1}-x_i) 
f_{\la_{N+1},\la_i}(x_{N+1}-x_i) \right]f_{\la_{N+1},\la_N} (x_{N+1}-x_N)}
\nonumber \\
&=&\prod_{i= 1}^{N-1} \left[f_{\la_{i+1},\la_i} (x_{i+1}-x_i)\right]
f_{\la_{N+1},\la_N} (x_{N+1}-x_N) \nonumber \\
&=&\prod_{i= 1}^N f_{\la_{i+1},\la_i} (x_{i+1}-x_i) \quad .
\eea
Referring back to Eq.(\ref{start}), we see that this proves the theorem.

\end{document}